\documentclass{article}
\usepackage[latin9]{inputenc}
\usepackage{geometry}
\usepackage{float}
\usepackage{amsmath}
\usepackage{amssymb}
\usepackage[unicode=true,pdfusetitle,
 bookmarks=true,bookmarksnumbered=false,bookmarksopen=false,
 breaklinks=false,pdfborder={0 0 0},backref=section,colorlinks=false]
 {hyperref}

\makeatletter

\providecommand{\\}{\\}

\usepackage{graphicx}
\usepackage{grffile}
\usepackage{breakurl}
\usepackage{caption}

\providecommand{\\}{\\}

\usepackage{fullpage}\usepackage[justification=centering]{caption}\usepackage[all]{hypcap}\usepackage{microtype}

\makeatother

\begin{document}

\title{SELECTION RULES FOR $^{48}$CR}

\author{Arun Kingan$^{1}$, Michael Quinonez$^{2}$, Xiaofei Yu$^{1}$ and Larry Zamick$^{1}$\\
\\
\textit{$^{1}$Department of Physics and Astronomy}, Rutgers University, Piscataway, New Jersey 08854
\\
\textit{$^{2}$Department of Physics and Astronomy}, Michigan State University, East Lansing, Michigan 48824}

\date{\today}
\maketitle
\begin{abstract}
In the single j shell (f$_{7/2}$)$^{48}$Cr is the first even-even
nucleus for which there are $T=0$ (isoscalar) $J=1^{+}$ states and
J=0 T=1 states. These states are studied here. This nucleus, in the
same model space, is midshell for both protons and neutrons. We can
assign a new quantum number to all these states which involves the
seniority of the protons and of the neutrons despite the fact that
the seniority itself is not a good quantum number. This then leads
to selection rules for electromagnetic transitions, e.g. B(M1), B(E2)
and to vanishings of static electric moments. 
\end{abstract}

\section{Introduction.}

In this work, we examine a nucleus which, in the single j shell model
space is at midshell. The nucleus is $^{48}$Cr, which can be viewed
as having 4 protons and 4 neutrons in the f$_{7/2}$ shell or as 4
proton holes and 4 neutron holes in that same shell. We will show
that there is a quantum number associated with all the states in this
model space, one which involves the seniority of the protons and of
the neutrons - this despite the fact that seniority itself is not
conserved. The good quantum number leads to selection rules for electromagnetic
transitions and static moments rules. In particular, we focus on B(M1)'s,
B(E2)'s and quadrupole moments of excited states. A second point of
interest is the fact that $^{48}$Cr is the first even-even nucleus
for which new states appear in a single j shell calculation - these
are J=1$^{+}$ T=0 and J=0$^{+}$ T=1. We also perform large space
calculations to see how things evolve.

Some of the points have been briefly presented in ref {[}1{]} but we
here add some additional observations e.g. of new J=0+ T=1 states,
vanishing quadruple moments, fluctuating quantum numbers along the
yrast band etc. Most important we now consider many large space
calculations to be compared with the simpler ones.

\section{New States in $^{48}$Cr in the single j shell model space.}

The nucleus $^{48}$Cr is of interest for a variety of reasons. Note
that in the single j shell (f$_{7/2}$) there are no $J=1^{+}$ $T=0$
in $^{44}$Ti. There are such states in the odd-odd nucleus $^{46}$V.
However, the first even-even nucleus for which there are isoscalar
J= 1$^{+}$ states in the single j shell configuration is $^{48}$Cr.
Likewise this is the first even-even nucleus for which there are J=0$^{+}$
T=1 states in the single j shell. We wish to study such states in
this section both in single j and the complete f-p space. 

Another point of interest is that in the single j shell we are at
midshell and this leads to selection rules which will be discussed
in the next section.

\section{Selection Rules for any N=Z Nucleus.}

Before discussing selection rules specific to the midshell nucleus
$^{48}$Cr let us consider such rules common to any N=Z nucleus.
Note that in this single j model space all B(M1)'s from $J=1^{+}$
$T=0$ states to $T=0$ states vanish for all $J_{f}$ (0, 1, 2).
This can be explained by the fact that in the limited model space
(f$_{7/2}$) the isoscalar M1 operator is proportional to $J=(J_{p}+J_{n})$
, total angular momentum operator. This operator, acting on a state
$\left|\alpha Jm\right\rangle $ will create a state in the same $\left|\alpha J\right\rangle $
multiplet and thus will not induce M1 transitions to different multiplets.
Also in this model space all transitions from 1$^{+}$ $T=1$ states
to other $T=1$ states ($J_{f}=$ 0, 1, 2) also vanish. This is a
known result which can be related to the vanishing, in an $N=Z$ nucleus,
of the Clebsch-Gordan coefficient (1 1 0 0 \textbar{}10).

\section{Selection Rules in for Midshell Nuclei: Calculations of B(M1)'s and
B(E2)'s in $^{48}$Cr.}

Some of the zeros, however, are specific to $^{48}$Cr. In the single
j shell we are at midshell. The 4 protons and 4 neutrons can also
be regarded as 4 proton holes and 4 neutron holes. As first noted
by Escuderos, Zamick {[}1{]} and Bayman {[}2{]} and shown analytically
by Neergaard {[}3{]}, the quantity $S=(-1)^{(v_{p}+v_{n})/2}$ is
a good quantum number, where $v_{p}$ and $v_{n}$ are the seniorities
of the protons and neutrons respectively. With the MBZE interaction
{[}2{]} the $J=0^{+}$ ground state has $S=+1$. With the same interaction
the yrast states of even $J$ have $S=(-1)^{J/2}$ i.e. $S=+1$ for
$J=0_{1}$, $S=-1$ for $J=2_{1}$, and $S=+1$ for $J=4_{1}$ etc.
Along the yrast chain the B(E2)'s are large and for these we have
$S_{f}=-S_{i}$. In the single j model space the B(E2)'s for transitions
in which $S_{f}=S_{i}$ will vanish.

Since for a static electric moment, e.g. Q$($$2^{+}$$)$, the ''transition''
matrix element is from a state to itself there is no change of S and
so this moment vanishes. However, with configuration mixing one gets
a static quadrupole moment which is close to the rotational value.
In ref {[}4{]} the calculated values are Q= - 35.42 e fm$^{4}$ and
B(E2,2$\rightarrow0)$=312.37 $e^{2}$fm$^{4}$ This is consistent with the
formulae of the a simple rotor model {[}5{]}:

\begin{minipage}[t]{1\columnwidth}%
Q={[}(3K$^{2}$-J(J+1){]}/{[}(J+1){*}((2J+3){]} Q$_{0}$\textbar{}\textbar{}
B(E2,K J$_{1}$$\rightarrow$K J$_{2}$) =5/(16$\pi)$e$^{2}$Q$_{0}^{2}$
\textless{}J$_{1}$$2$K 0\textbar{}J$_{2}$K\textgreater{}$^{2}$ %
\end{minipage}

The respective values of Q$_{0}$ are 123.97 and 125.30.

We show a brief example of the selection rules in Table I. We consider
B(E2)'S from the $J=0_{1}^{+}$ S=+1 state first two $J=2^{+}$states.
The $2^{+}{}_{1}$ state has $S=-1$ whilst the $2^{+}{}_{2}$ state
has, just like the $J=0_{1}^{+}$ ground state. We use the Shell model
Code NUSHELLX of B. A. Brown and W.D.M. Rae {[}6{]}.

\begin{table}[H]
\caption{B(E2)'s in the large and small spaces (.e$^{2}$ fm$^{4}$). }
\centering %
\begin{tabular}{ccc}
\hline 
0$_{1}$ To  & Large  & Single j\\
\hline 
\hline 
2$_{1}$  & 1225.5  & 452.6\\
2$_{2}$  & 2.367  & 0\\
\hline 
\end{tabular}
\end{table}

With regards to magnetic dipole transitions it is easy to show that
for B(M1) not to vanish a necessary condition is that $S_{f}=S_{i}$
i.e. $\Delta S=0$. This can be easily shown by examining the wave
functions of MBZE. {[}2{]}. When the M1 operator acts on a basis state
${[}J_{p},J_{n}{]}$ it creates a state with the same ${[}J_{p},J_{n}{]}$
(including any internal quantum numbers). We then overlap with the
final state. In the latter only the component with the same ${[}J_{p},J_{n}{]}$
will contribute. If $D(J_{p}, J_{n})$ for the initial state is non-zero
then the corresponding coefficient for the final state will be non
zero only if $S_{f}=S_{i}$.

We next take a casual look at magnetic dipole transitions and look
for selection rules for B(M1) values. With the MBZE {[}2{]} interaction,
the $S$ values for the first 3 $J=0^{+}$ $T=0$ states are +1,-1
and +1 respectively whilst the only 2 $J=0^{+}$ $T=1$ states both
have $S=-1$. For $J=0^{+}$ $T=2$ all 3 states have $S=+1$. Hence
the first and third $J=1^{+}$ $T=1$ states will connect with all
three but the second will not connect with any $J=0^{+}$ $T=2$ states.
For the first 3 $J=1^{+}$ $T=0$ states the $S$ values are -1,+1
and -1; for $J=1^{+}$ $T=1$ they are +1,-1, and +1; for $J=2^{+}$
$T=0$ they are -1,+1,+1 and finally , for $J=2^{+}$ they are -1, +1, -1.

For the lowest \char`\"{}special\char`\"{} $J=1^{+}$ $T=0$ state
which has $S=-1$ there will be no transitions in the single j shell
model space to any $T=0$ states, and there will be non-zero B(M1)'s
only to the 2 $J=0^{{+}}$ $T=1$ states, to the second $J=1^{+}$
$T=1$ state and to the first and third $J=2^{+}$ $T=1$ states.

The above selection rules can be obtained as easy generalizations
of results for particles of one kind, as described e.g. in R.D. Lawson's
book {[}7{]} and based on early work by G. Racah {[}8{]}. See especially
eq. 3.59 and the discussions that follow. It is there shown that the
matrix element of the $O^{\lambda}$ operator for particles is related
to that for holes by a phase factor $(-1)^{1+\lambda+(v-v')/2}$.
In that work $v$ refers to the seniority of particles of one kind.
We simply replace $v$ by $(v_{p}+v_{n})$ and $v'$ by $(v'_{p}+v'_{n})$.
In order to get a non-vanishing matrix element, the phase factor must
be positive. For B(M1) $\lambda=1$ and for B(E2) $\lambda=2$. This
explains the selection rules in a more formal way.

We next consider what happens in the complete f-p shell model space.
In Table II we show the large space results from various $J=1$ $T=0$
states to lowest and second $J=0^{+}$ and $J=2^{+}$ states and likewise
from various $J=1^{+}$ $T=1$ states. All B(M1)'s in the upper half
are $T=0$ to $T=0$ transitions and indeed they would have vanished
in the single j-shell calculations. In the lower half of Table II
we have $T=1$ to $T=0$ transitions and indeed the B(M1)'s are on
the whole much larger. Let is focus on the lowest J=1 T=1 to the lowest
J=0 T=0 transition. The value of B(M1) is 1.101. The orbital value
is 0.3046 and the spin value is 0.2475. The amplitudes add constructively
to give the total B(M1). When considered in reverse i.e. from 0 to
1 the B(M1) is 3 times as large and is often compared to the idealized
purely orbital scissors mode.

\begin{table}[H]
\centering \caption{B(M1)'s in a Large Space $(\mu_{N}^{2}$) }
\begin{tabular}{ccccc}
\hline 
$J=1_{n}^{+}\;T=0\rightarrow$  & Lowest $J=0^{+}$  & Second $J=0^{+}$  & Lowest $J=2^{+}$  & Second $J=2^{+}$\\
\hline 
\hline 
$n=1$  & 0.2003 E-3  & 0.2124 E-4  & 0.1095 E-4  & 0.6845E-4\\
2  & 0.1343 E-1  & 0.1334 E-3  & 0.5432 E-2  & 0.1288 E-3\\
3  & 0.5903 E-3  & 0.9063 E-5  & 0.5268 E-4  & 0.7698 E-4\\
4  & 0.5461 E-4  & 0.2338 E-2  & 0.4631 E-4  & 0.3361 E-2\\
5  & 0.7451E-5  & 0.1677 E-5  & 0.2297 E-3  & 0.1098 E-2\\
\hline 
$J=1_{n}^{+}\;T=1\rightarrow$  & Lowest $J=0^{+}$  & Second $J=0^{+}$  & Lowest $J=2^{+}$  & Second $J=2^{+}$\\
\hline 
\hline 
$n=1$  & 0.1101 E+1  & 0.1221 E-1  & 0.4665 E0  & 0.5316 E-1\\
2  & 0.6551 E0  & 0.1813 E0  & 0.3838 E0  & 0.4569 E-1\\
3  & 0.1570 E0  & 0,2142 E0  & 0.9775E-1  & 0.6353 E0\\
4  & 0.2353 E0  & 0.1709 E0  & 0.1768 E-2  & 0.4243 E0\\
5  & 0.5526 E-1  & 0.2574 E0  & 0.4835 E-2  & 0.2511 E0\\
\hline 
\end{tabular}
\end{table}

In Table III we show B(E2)'s from various $J=1^{+}$ $T=0$ states
to the 2 lowest $J=2^{+}$ and likewise from various $J=1^{+}$ $T=1$
states. The largest B(E2) in Table III is 25.32 e$^{2}$fm$^{4}$.
This is considerably smaller than value 1225.5 e$^{2}$fm$^{4}$
for the collective $J=0^{+}\rightarrow J=2^{+}$ transition shown
in Table I.

\begin{table}[H]
\centering \caption{B(E2)'s in a Large Space( e$^{2}$fm$^{4}$) . }
\begin{tabular}{ccc}
\hline 
$J=1_{n}^{+}\;T=0\rightarrow$  & Lowest $J=2^{+}$ State  & Second $J=2^{+}$ State\\
\hline 
\hline 
$n=1$  & 0.1897 E2  & 0.4232 E-2\\
2  & 0.5243 E1  & 0.5807 E0\\
3  & 0.8518 E-1  & 0.2532 E2\\
4  & 0.2944 E0  & 0.1551 E-1\\
5  & 0.8348 E-3  & 0.4874 E-1\\
\hline 
$J=1_{n}^{+}\;T=1\rightarrow$  & Lowest $J=2^{+}$ State  & Second $J=2^{+}$ State\\
\hline 
\hline 
$n=1$  & 0.6694 E1  & 0.1909 E-1\\
2  & 0.4376 E1  & 0.2359 E-1\\
3  & 0.3735 E0  & 0.2125 E1\\
4  & 0.1676 E-2  & 0.7543 E0\\
5  & 0.1546 E0  & 0.5553E-1\\
\hline 
\end{tabular}
\end{table}

\section{B(M1) Transitions Involving New States}

In Table IV we show M1 transitions between new states - that is states
which do not appear in the single j shell of even-even nuclei lighter
than 48Cr. We first note that the lowest J=0 T=1 state is at a rather
high excitation energy 8.89 MeV. The lowest J=1 T=0 state is 4.81
MeV. There are a few significant B(M1)'s e.g. from J=1T=0 at 4.184
MeV to three J=0 T=1 states at 10.20,10.53 and 10.78 with B(M1) values
0.235, 0.495 and 0.379

\begin{table}[H]
\centering \caption{$^{48}$Cr B(M1)'s from J=1, T=0 States to J=0, T=1 States (transitions
from columns to rows)}
\label{my-label} %
\begin{tabular}{c||c c c c}
\hline 
Excitation Energy  & 4.8143  & 6.5689  & 7.3052  & 7.975 \\
\hline 
\hline 
8.8901  & 0.106  & 0.1914  & 0.000748  & 0.01852 \\
9.044  & 0.003637  & 0.09062  & 0.000575  & 0.1085 \\
9.4744  & 0.0168  & 0.02171  & 0.004442  & 0.1682 \\
10.2023  & 0.2356  & 0.004279  & 0.1618  & 1.74E-06 \\
10.5339  & 0.4952  & 0.03943  & 0.01063  & 0.005871 \\
10.7829  & 0.3787  & 0.0461  & 0.04746  & 0.1424 \\
11.1399  & 0.01044  & 0.1266  & 0.02773  & 0.06704 \\
11.2304  & 0.142  & 0.04477  & 2.45E-07  & 0.01947 \\
11.7178  & 0.0617  & 0.1182  & 0.01249  & 0.1123 \\
11.89  & 0.1114  & 0.3067  & 0.01444  & 0.003353 \\
\hline 
\end{tabular}
\end{table}

\section{Closing Remarks : Relations to Other Nuclei }

Before closing we wish to make some comparisons other special nuclei
in the single j shell model space. In general the wave functions {[}2{]}
can be written as $\sum$D$^{I}$$^{T}$(J$_{p}$J$_{n}$) {[} J$_{p}$J$_{n}$){]}$^{I}$
where D$^{I}$$^{T}$(J$_{p}$J$_{n}$) is the probability amplitude
that in a state of total angular momentum I and isospin T the protons
couple to J$_{p}$ and the neutrons to J$_{n}$. We first consider
other N=Z nuclei such as $^{44}$ Ti. We note that the wave functions
have the property D(J$_{n}$,J$_{p}$) = (-1)$^{(I+T)}$D(J$_{p}$J$_{n}$).
Thus, for example we can see visually what states have isospin one.
for even J T=1 D(J,J)=0.

We next consider states with the same number of neutron holes as protons
e.g.$^{48}$Ti with 2 protons and 2 neutron holes. It was noted by
McCullen, Bayman and Zamick {[}9{]} and can be seen visually {[}2{]}
that D$^{I}$$^{T}$(J$_{p}$J$_{n}$)= (-1)$^{s}$ D$^{I}$$^{T}$(J$_{n}$J$_{p}$)
where the authors called ''s'' the signature quantum number. Unlike
the case of the N=Z nuclei states of different signature ca have the
same isospin.

Finally we come to midshell $^{48}$Cr. We here emphasize the visual
aspects. As shown in the appendix one can see a lot of zeros in the wave
functions that one does not see in other nuclei. Further analysis leads
us to the fact that (-1)$^{(v_{n}}$$^{+v_{p}}$$^{)}$$^{/2}$ is
a good quantum number. This is all the more remarkable since nether
v$_{p}$ or v$_{n}$ or v are good quantum numbers.

For future experiments, we mention in ref {[}10{]} calculations of
the scissors mode excitations in $^{48}$Cr. There should be large
M1 excitations of a 1$^{+}$ T=1 state in $^{48}$Cr. Since $^{48}$Cr is unstable one cannot directly excite
this state with electrons. However, recent experiments involving deexcitations, albeit
from a different nucleus from the scissors mode {[}10{]} to several
branches could be promising as applied to this nucleus.

\section{Acknowledgments}

M.I.Q. was supported by the REU program via an NSF grant PHY-1560077.
He is currently a student at Michigan State University.
A.K thanks the Rutgers Aresty Research Center for Undergraduates for
support during the 2016 summer session and the Richard J. Plano Summer
Research Internship for support during the 2017 summer session

\newpage{}

\section*{Appendix}

As an example we show the wave functions of the first 2 J=0$^{+}$
T=0 states in$^{48}$Cr from Ref {[}2{]}.

\begin{table}[H]
\centering \caption{Selected single j-shell wave functions with the MBZE interaction {[}2{]} in $^{48}$Cr.}
\begin{tabular}{c c c}
\hline 
Energy (MeV)  & 0.0000  & 5.5785\\
\hline
J$_{P}$J$_{n}$  & J=0$^{+}$  & J=0$^{+}$ \\
\hline 
\hline 
00  & 0.7494  & 0\\
22  & 0.5445  & 0\\
22{*}  & 0  & 0.6738\\
2{*}2  & 0  & 0.6738\\
2{*}2{*}  & 0.1243  & 0\\
44  & 0.1951  & 0\\
44{*}  & 0  & 0.2144\\
4{*}4  & 0  & 0.2144\\
4{*}4{*}  & 0.2521  & 0 \\
5{*}5  & 0.0932  & 0 \\
66  & 0.1231  & 0 \\
8{*}8{*}  & 0.0393  & 0 \\
\hline 
\end{tabular}
\end{table}

In Table V the {*} designates a seniority 4 basis state. The wave
functions of MBZE are written as follows: 
\begin{equation}
\Psi=\sum D^{\alpha J}(J_{p},J_{n}){[}J_{p},J_{n}{]}^{J}
\end{equation}
The value e.g. 0.5445 is the probability amplitude that in the lowest
J=0$^{+}$state.the 2 protons couple to angular momentum 2 and likewise
the 2 neutrons. The lowest J=0$^{+}$state has S=+1 and the next one
has S=-1. From ref {[}2{]} we see that there are 4 J=0$^{+}$ T=0
S=+1 states and 2 J=0$^{+}$ T=0 S=-1 states. This is true for any
charge independant interaction. For J=1$^{+}$ T=0 all states have
S=-1.

\begin{thebibliography}{10}
\bibitem[1]{} A.Escuderos and L.Zamick ,Romanian Journal of Physics,
Vol. 58, Nos. 9-10, pp 1064-1075, (2013)

\bibitem[2]{} A.Escuderos, L.Zamick and B.F. Bayman, arXiv:nucl-th/0506050
(2006)

\bibitem[3]{} K. Neergaard , Phys. Rev. C 91, 044313 (2015)

\bibitem[4]{} L. Zamick, Y.Y. Sharon,J.S.Q. Robinson and M. Harper,Phys.
Rev.C91,004321 (2015)

\bibitem[5]{} A. Bohr and B.R. Mottelson, NUCLEAR STRUCTURE Vol.
II,W.A.Benjamin Inc, Reading Massachusetts,(1975)

\bibitem[6]{} The Shell Model Code NUSHELLX@MSU, B.A. Brown
and W.D.M. Rae , http://www.sciencedirect.com/science/article/pii/S0090375214004748

\bibitem[7]{} R.D. Lawson,Theory Of The Nuclear Shell Model,
Clarendon Press, Oxford (1980)

\bibitem[8]{} G. Racah,Phys. Rev. 63, 367 (1943).

\bibitem[9]{} J.D. McCullen, B.F. Bayman and L. Zamick, Phys.
Rev., B515, 134 (1964)

\bibitem[10]{} S. J. Q. Robinson, L. Zamick, A. Escuderos, R.
W. Fearick, P. von Neumann-Cosel, A. Richter, Phys.Rev. C73 (2006)
037306

\end{thebibliography}
\end{document}